\documentclass{aa}  
\usepackage{graphicx}
\usepackage{txfonts}
\usepackage{color}
\usepackage[utf8]{inputenc}
\newcommand{\bhac}{\texttt{BHAC}~}

\newcommand{\eg}{e.g.,~}
\newcommand{\ie}{i.e.,~}

\usepackage{hyperref}
\hypersetup{
  citecolor=cyan,      
  filecolor=magenta,      
  urlcolor=magenta            
}

\begin{document} 

   \title{The impact of resistivity on the variability \\
   of black hole accretion flows}


   \author{Antonios Nathanail\thanks{anathanail@academyofathens.gr}
          \inst{1,2},
          Yosuke Mizuno\inst{3,4,2}, 
           Ioannis Contopoulos\inst{1}, 
           Christian M. Fromm\inst{5,2,6},  Alejandro Cruz-Osorio\inst{7}, \\
  Kotaro Moriyama\inst{8,2}  \and Luciano Rezzolla\inst{2,9,10}  }

   \institute{Research Center for Astronomy and Applied Mathematics,
 Academy of Athens, Athens 11527, Greece 
         \and
                 Institut f\"ur Theoretische Physik, Goethe Universit\"at
  Frankfurt, Max-von-Laue-Str.1, 60438 Frankfurt am Main, Germany 
  \and
    Tsung-Dao Lee Institute, Shanghai Jiao Tong University, Shanghai 201210, China \label{affil:shanghai}
  \and%
  School of Physics and Astronomy, Shanghai Jiao Tong University, Shanghai 200240, China \label{affil:shanghaicollege}
  \and%
 Institut f\"ur Theoretische Physik und Astrophysik,
  Universit\"at W\"urzburg, Emil-Fischer-Strasse 31, 97074 W\"urzburg,
  Germany 
  \and
        Max-Planck-Institut f\"ur Radioastronomie, Auf dem H\"ugel 69,
  D-53121 Bonn, Germany 
  \and 
  Instituto de Astronom\'{\i}a, Universidad Nacional Aut\'onoma de M\'exico, AP 70-264, 04510 Ciudad de M\'exico, Mexico
  \and  
  Instituto de Astrofísica de Andalucía, Gta. de la Astronomía, s/n, Genil, 18008 Granada
  \and
        School of Mathematics, Trinity College, Dublin 2, Ireland 
  \and
        Frankfurt Institute for Advanced Studies, Ruth-Moufang-Str. 1,
  60438 Frankfurt am Main, Germany
             }

   \date{Received July 15, 2024; Accepted November 19, 2024}

 
  \abstract
   {The accretion of magnetized plasma onto black holes 
   is a complex and dynamic process, where the magnetic 
   field plays a crucial role. The amount of magnetic 
   flux accumulated near the event horizon significantly 
   impacts the accretion flow behavior. Resistivity, a 
   measure of how easily magnetic fields can dissipate, 
   is thought to be a key factor influencing this process.}
   {This work explores the influence of resistivity 
   on accretion flow variability. We investigate
   simulations reaching the magnetically 
   arrested disk (MAD) limit and those with an initial 
   multi-loop magnetic field configuration.}
   {We employ 3D resistive general relativistic 
   magnetohydrodynamic (GRMHD) simulations to 
   model the accretion process under various 
    regimes, where resistivity has a global uniform value. }
   {Our findings reveal distinct flow behaviors 
   depending on resistivity. High resistivity 
   simulations never achieve the MAD state, 
   indicating a disturbed magnetic flux accumulation 
   process. Conversely, low resistivity simulations 
   converge towards the ideal MHD limit. The
   key results are:
i) For the standard MAD model, resistivity plays a 
minimal role in flow variability, suggesting that 
flux eruption events dominate the dynamics.
ii) High resistivity simulations exhibit strong 
magnetic field diffusion into the disk, rearranging  
efficient magnetic flux accumulation from the 
accretion flow.
iii) In multi-loop simulations, resistivity 
significantly reduces flow variability, which was not expected. However, 
magnetic flux accumulation becomes more variable 
due to frequent reconnection events at very low resistivity values.}
   {This study shows that resistivity affects 
   how much the flow is distorted due to magnetic field 
   dissipation. Our findings provide new 
   insights into the interplay between magnetic 
   field accumulation, resistivity, variability 
   and the dynamics of black hole accretion}


\titlerunning{Variability of resistive black hole accretion}
\authorrunning{Nathanail et al.}
   \maketitle
%

\section{Introduction}

The accretion of plasma onto black holes is the basis for studying and
analyzing observations of black holes. More specifically, the
supermassive black hole, Sgr\,A*, located at the center of our galaxy,
together with M87* have served as  subjects for numerous  
multi-wavelength observation campaigns \citep{Falcke1998,  
 Baganoff2001, Genzel2003, Doeleman2008, Hada2013, Kim2018}. 
 The Event Horizon Telescope (EHT) Collaboration has recently achieved a remarkable milestone by capturing groundbreaking images of black holes, revealing a luminous ring surrounding a prominent black hole shadow \citep{EHT_M87_PaperI, EHT_SgrA_PaperI} and an ordered magnetic field favouring the magnetically arrested disk MAD state \citep{EHT_M87_PaperVIII, EHT_SgrA_PaperVIII_etal}.
These observations highlight the limitations of current theoretical models in explaining the variability observed in the light curves of these objects 
\citep{EHT_SgrA_PaperV}. Understanding and characterizing this 
variability is essential for interpreting both black hole images and light 
curve data \citep{Burke2021, Broderick2022, Georgiev2022, Satapathy2022}.

MAD models are commonly used to describe
active galactic nuclei (AGN) with jets \citep{Bisnovatyi1974B, 
Narayan2003,Osorio2022,Fromm2021b}. As accretion progresses, magnetic
flux accumulates near the black hole's event horizon. The resulting 
magnetic pressure eventually balances the ram pressure of the disk, 
reaching equipartition and significantly impeding further accretion 
\citep{Igumenshchev2003,Igumenshchev2008,Tchekhovskoy2011}. However, 
three-dimensional, non-axisymmetric processes, like the magnetic Rayleigh -
Taylor instability, allow for continued accretion \citep{Papadopoulos2019}. 

\begin{table*}
   \caption{Initial parameters for the models considered 
}
\label{table:models} 
  \centering
        \begin{tabular}{l|c|c|c|c|l|c|c|c|c|c|} \hline \hline
                model  &  \underline{$2p_{\rm max}$}  &
                $\sigma_{\rm max}$ & module
                &  $\eta$ & $N_r \times N_{\theta} \times N_{\phi}$
                & $r_{\rm in}$ & $r_{\rm p_{\rm max}}$ & $r_{\rm max}$
                &$\langle Q_{\theta}\rangle$ & $\langle Q_{\theta}\rangle$\\
                &$(B^2)_{\rm max}$&$\times10^{-4}$&&$5\times$&&&&&$<4000\,{\rm M}$&$>4000\,{\rm M}$\\
                \hline
         ${\rm MAD.S.100.E.00}$ & $100$   &  $2$   & Ideal      & 0 & $384 \times 192 \times 192$ 
        &20&41&2500&>10&>10\\
        ${\rm MAD.S.100.E.-6}$   & $100$    &  $2$   & Resistive & $10^{-6}$& $384 \times 192 \times 192$
        &20&41&2500&>10&~3\\
        ${\rm MAD.S.26.E.-6}$   & $26$    &  $4$   & Resistive & $10^{-6}$& $384 \times 192 \times 192$
        &20&41&2500&>10&6\\
        ${\rm MAD.S.26.E.-5}$   & $26$    &  $4$   & Resistive & $10^{-5}$& $384 \times 192 \times 192$
        &20&41&2500&>10&6\\
        ${\rm MAD.S.26.E.-5.HR}$   & $26$    &  $4$   & Resistive & $10^{-5}$& $1050 \times 768 \times 384$
        &20&41&500&>10&>10\\
        ${\rm MAD.S.26.E.-4}$  & $26$    &  $4$   & Resistive & $10^{-4}$& $384 \times 192 \times 192$
        &20&41&2500&>10&>10\\
        ${\rm MAD.S.26.E.-3}$   & $26$    &  $4$   & Resistive & $10^{-3}$& $384 \times 192 \times 192$
        &20&41&2500&>10&~5\\
        
        \hline \hline
        
       ${\rm ML.S.26.E.00}$   & $26$   &  $2.5$   & Ideal & 0 & $1050 \times 768 \times 384$
        &6&12&500&>10&>10\\
       ${\rm ML.S.26.E.-5}$   & $26$   &  $2.5$   & Resistive & $10^{-5}$& $1050 \times 768 \times 384$
        &6&12&500&>10&>10\\
                \hline \hline
        \end{tabular} 
\tablefoot{
The first
        column refers to the type of accretion model, whereas the second column
        to the initial maximum values of $2p/B^2$ and the third column to the initial maximum magnetization $\sigma_{\rm max}$.
        The fourth column corresponds to the
        module used in \bhac Ideal or Resistive and the next column to
        the value of the uniform resistivity  $\eta$, employed (Ideal models have
        $\eta = 0$). The next column reports the resolution of each run, whereas columns 7 and 8 report the inner
radius and maximum density radii of the torus, and the 9th column the outer
radius of the domain, in units of $r_g$. The two last columns report the average MRI quality
factor $Q_{\theta}$ at the heart of the torus \citep{Porth2019_etal} for time before
(column 10) and after $4000\,{\rm M}$ (last column)\footnote{More details on the MRI quality factor can be found in Appendix C.}.
}
 \end{table*}
Resistivity plays a
crucial role, by influencing the magnetic reconnection process itself. The angular momentum transport and the amplitude of magnetic energy after saturation can be  significantly reduced by finite resistivity, as
the Magneto-rotational Instability (MRI) can be influenced under certain conditions in the disk, which would affect the angular momentum transport 
 \citep{Pandey2012}.
A common approach is to introduce a global 
constant value for resistivity in relativistic MHD simulations \citep{DelZanna2016,
Ripperda2019, Mattia2023}. 
However, a more realistic and physically motivated model for 
resistivity 
is needed, 
considering its non-uniform nature. Ideally, resistivity should be 
significant only on local scales, like X-points where reconnection 
occurs, while leaving the global dynamics largely unaffected 
(\citep{Selvi2023}). 
A similar investigation took place a decade ago in pulsars  
but soon was replaced by Particle-in-cell simulations
\citep{LiJ2012, Kalapotharakos2012b}.


This study investigates the impact of resistivity on the dynamics of 
the accretion flow using a global prescription to describe 
this effect. We present results of 3D GRMHD simulations for resistive
MAD models and also accretion models with an initial multi-loop 
magnetic field configuration, in which there is no steady magnetized 
funnel above the black hole but reconnecting current sheets are 
periodically formed in this region \citep{Nathanail2020, 
Chashkina2021, Nathanail2021b}. The latter scenario may be 
specifically relevant for modeling the future observations of the galactic center, Sgr\,A* \citep{Nathanail2022}. 
It is important to note that resistivity was thought not 
 to have a significant impact on the 
variability of the accretion process, let alone reduce it. 
However, a key  and unexpected result of our study 
is that the inclusion of resistivity does influence and even lower the variability for a specific  accretion model,
the multi-loop one.

The paper is structured as follows: In Section \ref{sec:main} we
present our simulations, with subsection \ref{sec:setup}
focusing on the numerical setup and subsection \ref{sec:acc}
reporting the details of the accretion process. In subsection \ref{sec:var}
we analyze and discuss the variability of the accretion flow, 
and in Section \ref{sec:con} we present our conclusions.

\section{Resistive GRMHD simulations}
\label{sec:main}

%
\subsection{Numerical setup}
\label{sec:setup}
The numerical configuration comprises a Kerr black hole spacetime and an
initially perturbed torus infused with a poloidal magnetic field. All
simulations conducted in this study are performed in three spatial
dimensions, employing the GRMHD code \bhac \citep{Porth2017}. This code 
utilizes second-order shock-capturing finite-volume methods and has been 
extensively utilized in various investigations \citep{Nathanail2018c, 
Mizuno2018, Nathanail2020b}. It employs the constrained-transport method 
\citep{DelZanna2007} to ensure a divergence-free magnetic field
\citep{Olivares2019} and has undergone comprehensive testing and
comparisons with similar-capability GRMHD codes \citep{Porth2019}.

We explore two sets of models, MAD and multi-loop configurations. For
the first, the initial data consist of an equilibrium torus with a constant 
specific angular momentum of $\ell=6.76$ 
\citep{Fishbone76} orbiting around 
a Kerr black hole with dimensionless spins of $a = 0.937$. 
The magnetic field is initialised as a nested loop described by the vector
potential:
\begin{align}
        A_{\phi}=\max \left(\left(\frac{\rho}{\rho_{\max}}\left( \frac{r}{r_{\rm in}} \right)^3
        {\rm sin^3} \theta \exp \left( \frac{-r}{400} \right)\right)-0.2,0\right),
\label{Aphi}
\end{align}
where the maximum rest-mass density in the torus is denoted with
$\rho_{\rm max}$.

For the multi-loop models for a spin of $a=0.5$ and $\ell=4.28$, the initial magnetic field consists of a series of nested loops with
alternating polarity.
The  vector potential has the form:
\begin{align}
  &A_{\phi}\propto \max\left(\frac{\rho}{\rho_{\rm max}}-0.2,0\right)\cos((N
  -1)\theta) \sin\left(\frac{2\pi \, (r-r_{\rm in} )}{\lambda_r}\right),
\end{align}
\begin{figure*}
  \begin{center}
    \includegraphics[width=0.45\textwidth]{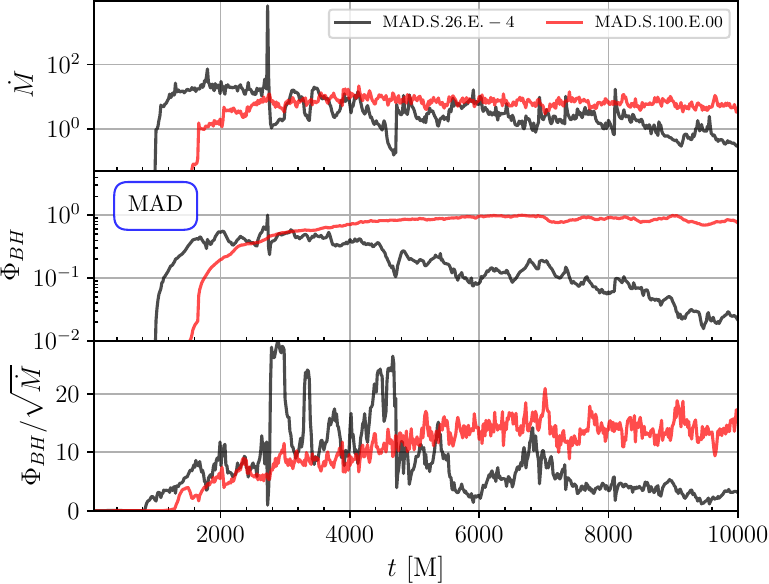}
    \includegraphics[width=0.45\textwidth]{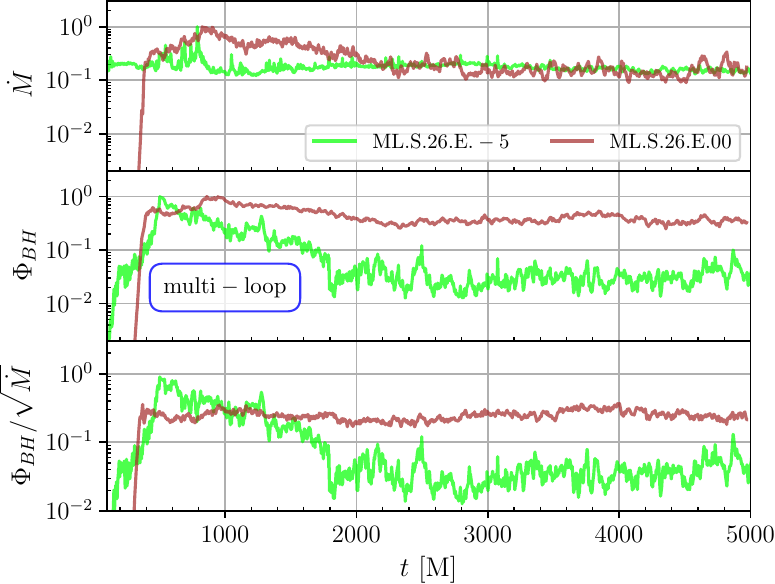}
  \end{center}
\caption{Upper panels: mass accretion rate, $\dot{M}$, through the black-hole
horizon, Middle panels: the magnetic flux accumulated on the black-hole
horizon, $\Phi_{\rm BH}$, Lower panels: the normalized magnetic flux accumulated on the black-hole
horizon, $\phi_{\rm BH}$. Left panels:  MAD models. Right panels: multi-loop models in both panels ideal and resistive (see Table \ref{table:models} for details.)%
}
    \label{fig:mdot}
\end{figure*}
\begin{figure}
  \begin{center}
    \includegraphics[width=0.45\textwidth]{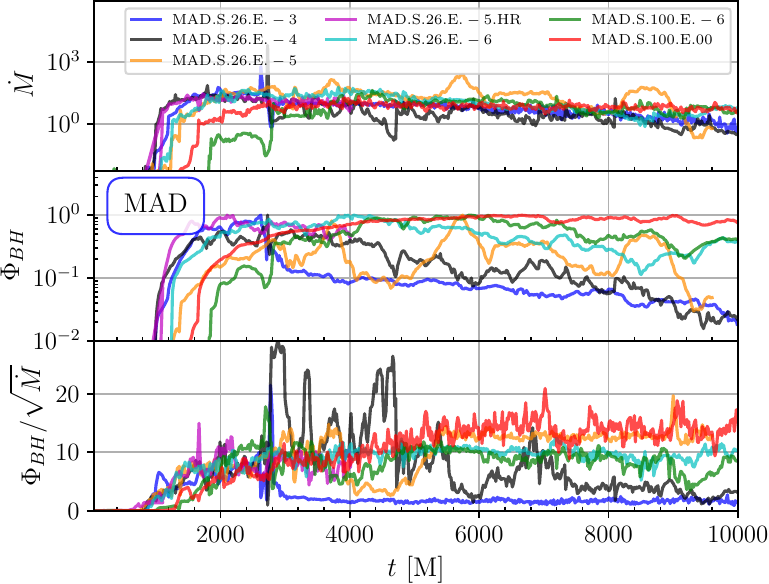}
  \end{center}
\caption{Same as Fig. \ref{fig:mdot} for all MAD models
(see Table \ref{table:models} for details.) 
}
    \label{fig:mdot-all}
\end{figure}

the additional parameters ($N=3$ and
$\lambda_r=2$) set the number and the characteristic length-scale of the
initial magnetic loops in the torus. 
presented and analysed in 2D and 3D \citep{Parfrey2015, Yuan2019, Yuan2019b, 
Mahlmann2020}. Two-temperature GRMHD simulations with a multi-loop magnetic 
field have also been conducted to trace electron heating through turbulence 
and reconnection, with findings suggesting that the electrons are often 
trapped in plasmoids \citep{Jiang2023}. Additionally, these simulations 
investigated the emission properties of the plasmoids \citep{Jiang2024}.

The computational domain adopts a spherical logarithmic Kerr-Schild
coordinate system.
In Table \ref{table:models} we report all the 3D simulations conducted
in this study. The initial field strength is set by the value of $2p_{\rm 
max}/(B^2)_{\rm max}$, where the location of maximum of fluid and magnetic pressure may 
not coincide. The ideal MAD model, ${\rm MAD.S.100.E.00}$,  
is the standard MAD model, typical in the 
literature , for comparison the same model is 
run with resistivity  $\eta = 5\times 10^{-6}$, ${\rm MAD.S.100.E.-6}$. 
To study the effect of magnetic field 
dissipation and its impact on the dynamics of the flow we 
increase the field strength for the rest of the models which results 
in a factor of $2$ larger initial maximum magnetization. 
The effective resolution is reported in column 6 of Table \ref{table:models}. 
Models ${\rm MAD.S.26.E.-5.HR}$, ${\rm ML.S.26.E.00}$ 
and ${\rm ML.S.26.E.-5}$ that have the 
highest resolution, end at $t=5000$ M.

For the simulations utilizing physical resistivity, i.e., $\eta\neq0$, we employ the
resistive GRMHD equations implemented in \texttt{BHAC}
\citep{Ripperda2019}. In this setup, we assume a uniform and
constant resistivity with a varying value of $\eta =5\times
10^{-3}- 10^{-6}$\footnote{For Sgr\,A*, and a magnetic field of $B\approx 30$ G, resistivity is $\eta=\eta_{\rm code}\times t_g$ s, thus $\eta =5\times 10^{-6}\times t_{g,SgrA\,*}=10^{-4}$ s in Gaussian/ESU units, for reference this value is well above the Spitzer resistivity ($\eta_{SP}=1.15\times 10^{-14}\,Z\,\ln{\Lambda}(T)^{-3/2}\approx 10^{-20}$ s, with $Z=1$ for hydrogen, $\ln{\Lambda}$ the Coulomb logarithm and $T$ the electron temperature) of the local plasma, but closer to the expected anomalous resistivity expected in astrophysical accretion disks}. The choice of a small resistivity, $\eta
=5\times 10^{-5}$ and $\eta =5\times 10^{-6}$
is deliberate, as it allows us to replicate
nearly ideal conditions for the accretion flow dynamics while
also permitting physical magnetic reconnection processes that may lead to the
formation of plasmoids \citep{Ripperda2019b, Ripperda2020}.

In jets of supermassive black holes, considering the gravitational radius, $L\approx r_g$, as the characteristic
length scale, resistivity of $\eta =5\times 10^{-6}$ yields 
a high Lundquist number of $S := Lu_A/\eta=r_g c/\eta > 10^4$, magnetization is large and the Alfv\`{e}n speed approaches the speed of light $u_A \approx c$ \citep{Guo2015}. This value
represents the minimum Lundquist number required to generate
plasmoids under the physical conditions examined in this study
\citep{Bhattacharjee2009, Uzdensky2010, Ripperda2019b,
Ripperda2020}. However, we also employ values of resistivity
higher than the minimum value to explore its impact 
on the properties of the accretion flow and the induced variability.

In the ideal-GRMHD models (refer to Table \ref{table:models}), the
dissipation of magnetic energy is solely a numerical effect.
However, previous 2D simulations have demonstrated that, increasing
the resolution still allows us to obtain physically meaningful
results, allowing a more detailed study of the solution's behavior \citep{Obergaulinger:2009, Rembiasz2017, Nathanail2020,
Obergaulinger2020}.

\begin{figure*}
  \begin{center}
    \includegraphics[width=0.97\columnwidth]{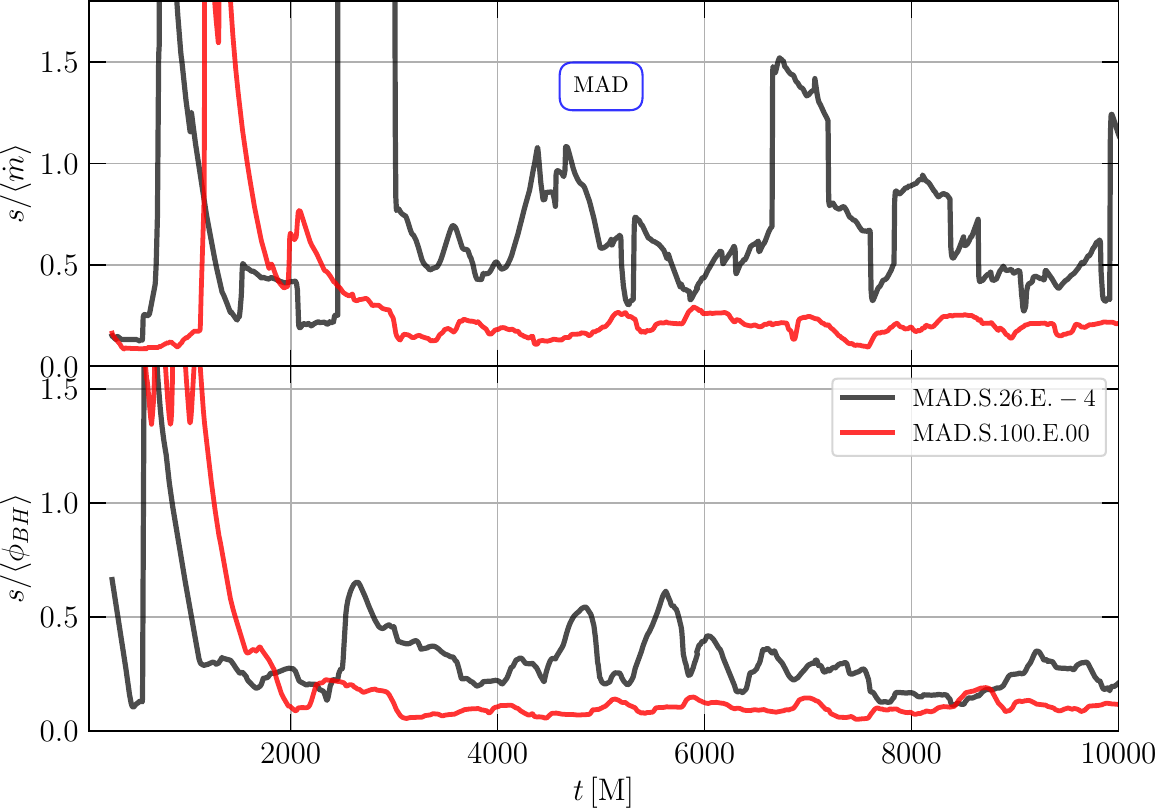}
    \includegraphics[width=0.96\columnwidth]{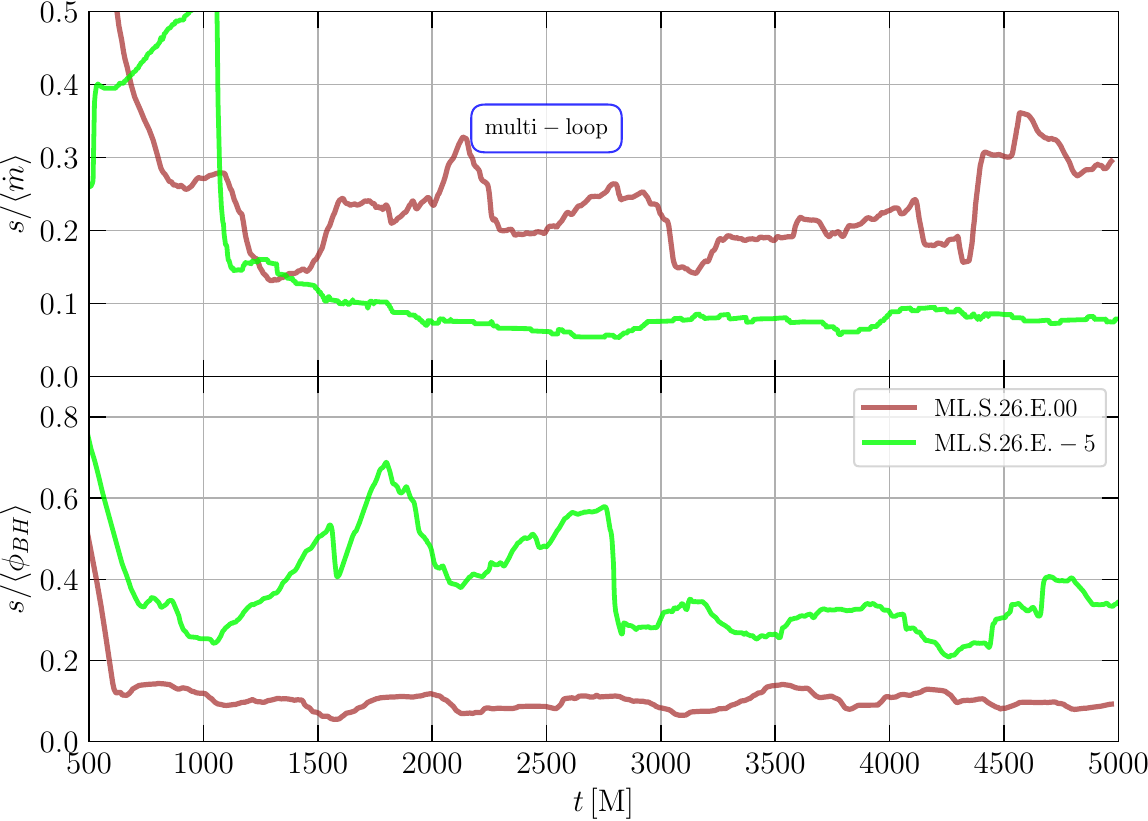}
  \end{center}
  \caption{ Upper panels: the measure of variability for the mass accretion rate
$s/\langle\dot{m}\rangle$, Lower panels: the measure of variability for the normalized magnetic flux accumulated on the black-hole horizon $s/\langle\phi_{BH}\rangle$, both for a time window of $\pm 270$ M
\citep{EHT_SgrA_PaperV} . Left panels: All MAD models ideal and with different resistivity, Right panels: multi-loop models ideal and resistive (see Table \ref{table:models} for details.) 
}
    \label{fig:var}
\end{figure*}
\begin{figure}
  \begin{center}
    \includegraphics[width=0.435\textwidth]{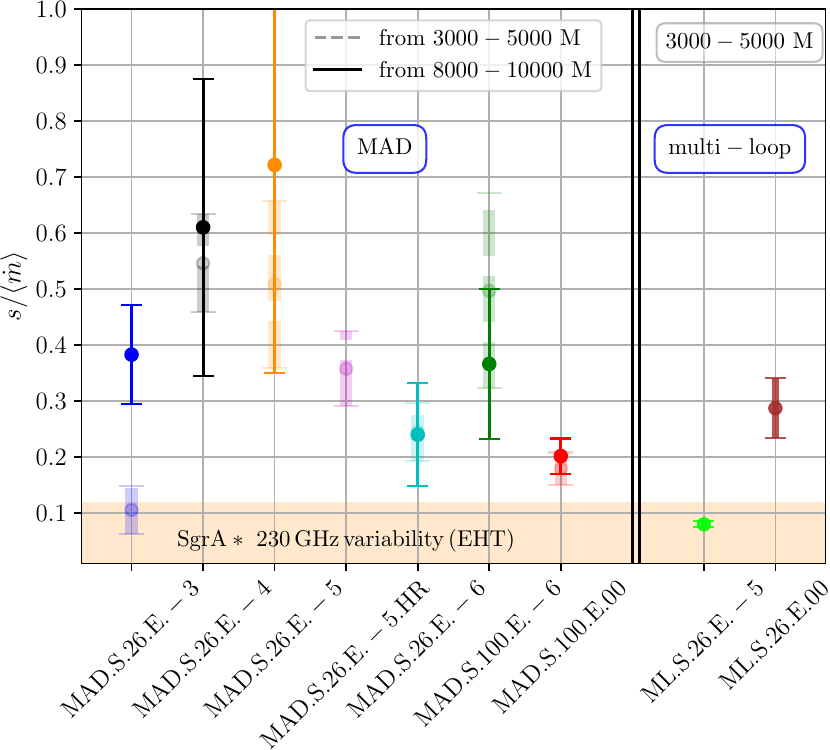}
  \end{center}
\caption{The measure of variability for the mass accretion rate
$s/\langle\dot{m}\rangle$, (error bars ~ $\pm 1s$) for  all MAD and multi-loop models in time windows $3000-5000$ M (dashed error bars) and $8000-10000$ M (solid error bars) respectively. 
}
    \label{fig:mean_var}
\end{figure}
\subsection{Properties of the accretion flow}
\label{sec:acc}

Our work primarily focuses on a single aspect, investigating the
influence of resistivity on the variability of the accretion flow.
To analyze the main properties of the accretion dynamics
we define the rest-mas accretion rate and the magnetic flux across the
horizon. The former is measured as:
\begin{align}
 \dot{M}:=\int_0^{2\pi}\int^\pi_0\rho u^r\sqrt{-g}\,d\theta d\phi\,,
\label{mdot}
\end{align}
where $\rho$ is the rest-mas density, $u^r$ is the radial component of the four-velocity and $\sqrt{-g}$ is the determinant of the spacetime metric. 
Its behaviour is reported as a function of time in the upper panels of 
Figs.~\ref{fig:mdot} and ~\ref{fig:mdot-all}. The left panels refer to 
two MAD models, whereas the right ones to the multi-loop models.
The magnetic flux accreted across the event horizon is defined as: 
\begin{align}
  \Phi_{\rm BH}:=\frac{1}{2}\int_0^{2\pi}\int^\pi_0 |B^r|\sqrt{-g}d\theta d\phi\,,
\label{phiBH}
\end{align}
while the ``normalized'' magnetic flux is defined as $\phi_{\rm BH}:= 
\Phi_{\rm BH}/ \sqrt{\dot{M}}$. In the middle (and lower) panels of Figs.~\ref{fig:mdot} and ~\ref{fig:mdot-all} the magnetic flux (and normalized magnetic flux 
respectively) is shown for all of the models considered.
The limiting normalized magnetic flux quoted by 
\citet{Tchekhovskoy2011} is $\phi_{\rm BH}=\phi_{\rm max}\approx 50$. In 
our simulations using Heaviside-Lorentz units (as opposed to Gaussian 
units) this value should be divided by a $\sqrt{4\pi}$, thus $\phi_{\rm 
BH}=\phi_{\rm max}\approx 50/\sqrt{4\pi}\approx 14$.

Accretion models with low resistivity ($\eta =5\times 10^{-5}$) exhibit 
a behavior consistent with standard MAD models.  
However, as resistivity increases ($\eta =5\times 10^{-4}$), 
the dynamics change. At $\eta =5\times 10^{-4}$, 
intense dissipation at the boundary of the magnetized funnel 
triggers a flare event lasting nearly $1000$ M (left panels 
of Fig~\ref{fig:mdot}). This event leads to a sharp decrease 
in magnetic flux at the horizon. The simulation eventually 
settles into a state with a much lower flux level. A complete 
discussion of magnetic field accumulation in resistive MAD models
can be found in Appendix A.

\subsection{Variability of the accretion flow}
\label{sec:var}

To measure the variability of the accretion flow we define the mean and its variance for the quantities introduced before, namely the mass accretion rate and the normalized magnetic flux. 
The computation is done at a specific point in the time series for a window of $\pm \, 270$ M\footnote{The time window was chosen to cover 3 hours of observational data for SgrA* see \citep{EHT_SgrA_PaperV} for details.}, thus defined as follows: 
\begin{align}
  \mu = \frac{1}{2n}\sum^{+n}_{i=-n} k_{i},  \qquad
  s^2 = \frac{1}{2n}\sum^{+n}_{i=-n} (k_i - \mu)^2,
  \label{eqvar}
\end{align}

where $n=270$ and $k_i$ is  the quantity under investigation, e.g., $\dot{M}$ or $\phi_{\rm BH}$.
Finally, we report the value of $s/\mu$, were $\mu = \langle\dot{m}\rangle$ or $ \langle\phi_{BH}\rangle$, which measures the variability for each of these quantities. In Fig.~\ref{fig:var}, we present the results of this procedure for the mass accretion rate (upper panels) and the normalized flux (lower panels) for two MAD models\footnote{Variability for all MAD models is shown in the Appendix B.} (in the left panel) and the multi-loop models (in the right panel). 

The variability imprinted in the normalized magnetic flux is similar in the MAD models (left panel). More specifically, at late times ($8000-10000$ M) it stabilizes around $~0.2$. Minimum variability occurs in the MAD model with $\eta =5\times 10^{-5}$. Intense reconnection near the horizon due to colliding flux tubes of opposite polarity significantly increases the variability of the normalized flux in the multi-loop models \citep{Nathanail2021b}. 

Mass accretion rate, a measure of how much mass is falling on the black hole, plays a crucial role in shaping the radiation light-curve, particularly at $230$ GHz. 
Studies have shown a close link between the variability of mass accretion rate and the variability of the light-curve at $230$ GHz \citep{Porth2019, Chatterjee2021}.  
For this reason we primarily focus 
on the variability of the mass accretion rate.

The results of the mean variability are summarised in Fig. \ref{fig:mean_var} for two different time windows, namely $3000-5000$ M and $8000-10000$ M. Variability is significantly reduced by the inclusion of resistivity only for the multi-loop model. 
For models capable of reaching the MAD state, the variability is not influenced significantly by the inclusion of resistivity. 
However, there is no simple relation between the variability of the flow and the resistivity of the plasma, as seen in Fig. \ref{fig:mean_var}.
A proper analysis of the $230$ GHz light curve, which is the EHT's target frequency, will be conducted in future studies.

\section{Conclusions}
\label{sec:con}
The key findings are the following:
   \begin{enumerate}
      \item For MAD models, resistivity of 
      $\eta =5\times 10^{-5}- 10^{-6}$ has minimal impact on the variability of the flow, indicating that the dynamics are primarily driven by magnetic flux eruption events.
      \item For MAD models, simulations with resistivity of 
      $\eta =5\times 10^{-3}- 10^{-4}$, show significant magnetic field diffusion into the disk, hindering the efficient accumulation of magnetic flux from the accretion flow.
      \item For multi-loop models, even if frequent reconnection events lead to increased variability in magnetic flux accumulation, resistivity significantly reduces the variability of the accretion flow. We need here to stress out that resistivity was not expected to have such an impact.
   \end{enumerate}

\begin{acknowledgements}
   Support comes from the ERC Advanced Grant ``JETSET:
Launching, propagation and emission of relativistic jets from binary
mergers and across mass scales'' (Grant No. 884631). 
YM is supported by the National Key R\&D Program of China (grant no. 2023YFE0101200), the National Natural Science Foundation of China (grant no. 12273022), and the Shanghai municipality orientation program of basic research for international scientists (grant no. 22JC1410600).
CMF is supported by the DFG research grant ``Jet physics on horizon scales and beyond" (Grant No.  443220636) within the DFG research unit ``Relativistic Jets in Active Galaxies" (FOR 5195). 
ACO gratefully acknowledges ``Ciencia Básica y de Frontera 2023-2024" program of the ``Consejo Nacional de Humanidades, Ciencias y Tecnología" (CONAHCYT, Mexico) project CBF2023-2024-1102 and SNI 257435.
This work was supported by computational time granted from the National Infrastructures for Research and Technology S.A. (GRNET S.A.) in the National HPC facility - ARIS - under project ID 16033. Simulations
were performed also on SuperMUC at LRZ in Garching, on the GOETHE-HLR cluster
at CSC in Frankfurt, and on the HPE Apollo Hawk at the High Performance
Computing Center Stuttgart (HLRS).
\end{acknowledgements}
%
%
%
\noindent\textit{\textbf{Data Availability.~}}
The data underlying this article will be shared on reasonable request to the corresponding author.

\bibliographystyle{aa}
\bibliography{aanda}

\newpage

\section*{Appendix A. Magnetic field accumulation in Resistive MAD simulations}
\label{appA}

\begin{figure}[b]
  \begin{center}
    \includegraphics[width=0.455\textwidth]{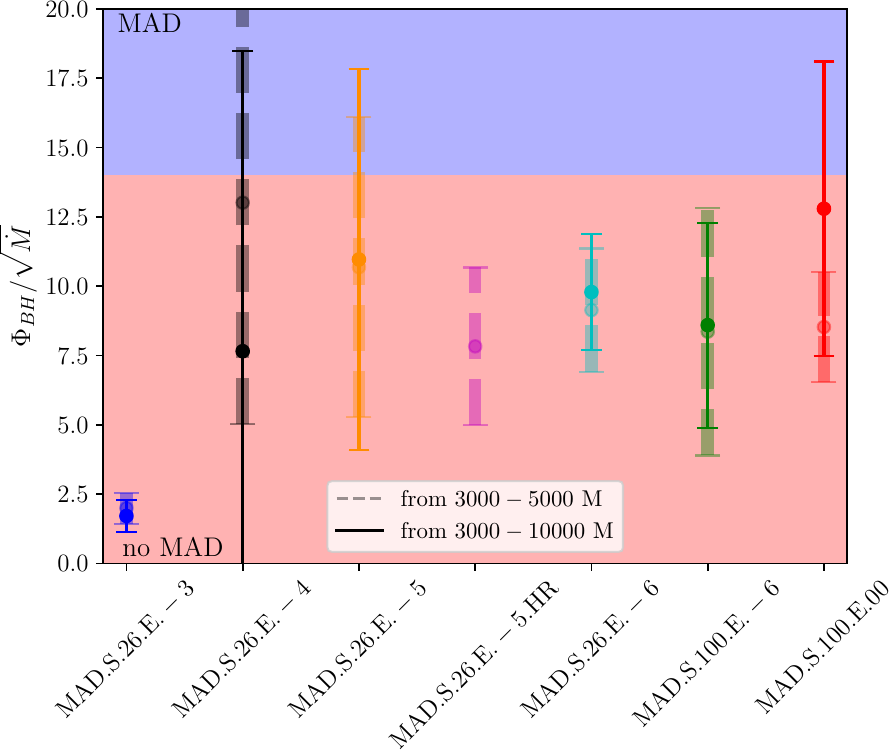}
  \end{center}
\caption{The mean normalized magnetic flux accumulated on the 
black-hole horizon (error bars indicate $\pm 1s$) for all MAD models in time windows $3000-5000$ M (dashed error bars) and $3000-10000$ M (solid error bars) respectively. The dashed horizontal line depicts the MAD saturation value $\phi_{\rm max}\approx 50/\sqrt{4\pi}\approx 14$. }
    \label{fig:mean}
\end{figure}

To better understand the dependence of simulation results on resistivity, 
Fig.~\ref{fig:mean} shows the average normalized magnetic flux on the 
horizon for all MAD models. The averages are taken over two-time windows: 
$3000-5000$ M and $3000-10000$ M. The figure clearly shows the flaring 
event for  model ${\rm MAD.S.26.E.-4}$, where the flux fluctuates 
near the MAD limit in the first window and then drops slightly. 
Model ${\rm MAD.S.26.E.-3}$  shows a substantially 
lower flux already in the first window, indicating a clear deviation from 
typical MAD behavior. Longer simulations could highlight 

Resistivity significantly impacts magnetic flux accumulation in 
MAD simulations due to energy dissipation at the edge of
the magnetized funnel. Models with a rather 
low resistivity ($\eta =5\times 10^{-5}$) exhibit 
a behavior consistent with standard MAD models.  
At the highest resistivity run ${\rm MAD.S.26.E.-3}$  the 
system settles quickly into a state with very low magnetic 
flux on the horizon (see left, upper and bottom panels of Fig~\ref{fig:mdot-all}).

The effect of numerical resolution was compared in two simulations 
with $\eta =5\times 10^{-5}$, where the high-resolution run is 
indicated as HR, ${\rm MAD.S.26.E.-5.HR}$'. The higher resolution allows for a more 
accurate representation of small-scale physical processes, such as 
reconnection events at current sheets. As a consequence, the HR simulation exhibits more efficient magnetic energy dissipation near the 
edge of the funnel, resulting in a lower average accumulated flux 
compared to the standard resolution simulations. 
However, a definitive 
conclusion on the quantitative impact of resolution would require running 
the HR simulations for a longer time to ensure we capture the
behavior of the system till $10000$ M. In general, longer simulations could highlight whether our results remain consistent or change over extended time
evolution.

Models with $\eta =5\times 10^{-6}$, suggest that a stronger 
initial magnetic field (in cyan, initial $\beta=26$) allows the system to accumulate more flux faster and 
reach a state closer to the ideal MAD limit even in the presence of some 
resistivity.  This comment must be the same for all models, meaning 
that initialization with higher $\sigma$ will eventually bring more 
magnetic flux at the event horizon. 
The simulations reveal that resistivity plays a crucial role in magnetic 
flux accumulation. Lower resistivity allows for behavior consistent with 
the standard MAD model. 
However, as resistivity increases, dissipation 
processes become more prominent, leading to flare events and a 
significant reduction in the accumulated magnetic flux.

\section*{Appendix B. Variability for all MAD models}
\label{appB}

\begin{figure}[t]
  \begin{center}
    \includegraphics[width=0.93\columnwidth]{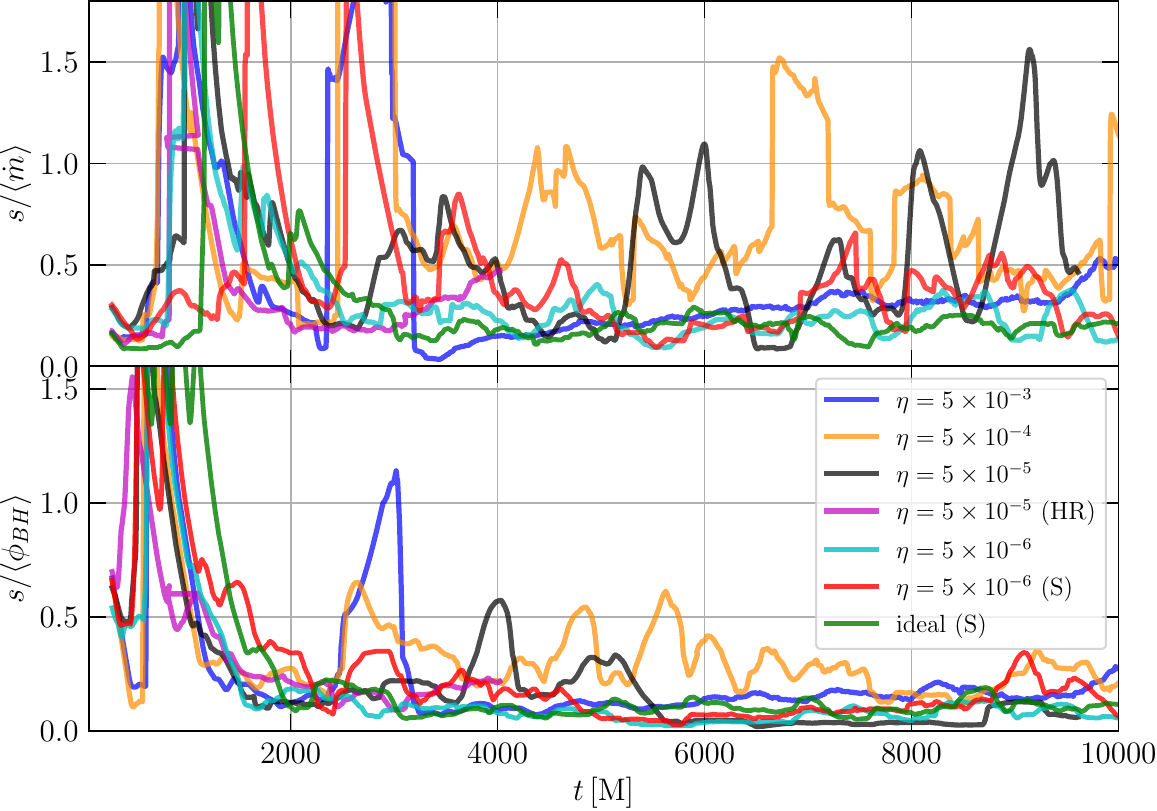}
  \end{center}
  \caption{ Upper panels: the measure of variability for the mass accretion rate
$s/\langle\dot{m}\rangle$, Lower panels: the measure of variability for the normalized magnetic flux accumulated on the black-hole horizon $s/\langle\phi_{BH}\rangle$, both for a time window of $\pm 270$ M
\citep{EHT_SgrA_PaperV}  for all MAD models ideal and with different resistivity (see Table \ref{table:models} for details.)}
    \label{fig:var-all}
\end{figure}

In this Appendix we provide details for the variability of all MAD models 
in our study. Fig.~\ref{fig:var-all} presents the results for 
$s/\langle\dot{m}\rangle$ (upper panel) and $s/\langle\phi_{BH}\rangle$ 
(lower panel) for a time window of $\pm 270$ M \citep{EHT_SgrA_PaperV}.

The model with the highest resistivity of $\eta =5\times 10^{-3}$ 
(${\rm MAD.S.26.E.-3}$, blue line), exhibits a large bump at early times 
(~$3000-3500$ M) likely corresponding to a rapid loss of magnetic flux. 
This is followed by a minimum variability state, which then steadily increases until the end of the simulation. Model ${\rm MAD.S.26.E.-4}$, which has a lower resistivity ($\eta =5\times 10^{-4}$, orange line) shows a similar initial variability bump and continues to exhibit recurring bumps every $1000 - 2000$ M.

As is seen in models ${\rm MAD.S.26.E.-5}$ and ${\rm MAD.S.26.E.-5.HR}$, higher resolution reduces the numerical diffusion and let only the physical resistivity to act.
Another point to make for the HR run is that it reduces slightly the variability in the 
first window, this was expected since having HR will impact any variability imposed 
from reconnection events.
Further reducing the resistivity to $\eta =5\times 10^{-6}$, results in a similar level of variability to the standard ideal MAD model. 
\bigskip

\section*{Appendix C. Variability of the Power of the jet}
\label{appC}

In this appendix we explore the variability on the jet power. For the 
multi-loop models such discussion does not make much sense, since it 
has been shown that there is no production of a steady 
jet\citep{Nathanail2020}. Multi-loop models exhibit periodic outbursts 
either from the upper or the lower hemisphere. Thus, the variability 
of the jet power will be discussed only for MAD models. 

The power  is measured through the energy flux that passes
through a 2-sphere placed at $50 \, r_g$. It is defined  as follows:
\begin{align}
  P_{\rm jet}:=\int_0^{2\pi}\int^\pi_0 (-T^r_t -\rho u^r)\sqrt{-g}d\theta
  d\phi\,,
\label{pjet}
\end{align}
where the integrand in \eqref{pjet} is set to zero  everywhere on the
integrating surface where $\sigma \leq 1$, in order to account only 
for the jet component. 

\begin{figure}[t]
  \begin{center}
    \includegraphics[width=0.96\columnwidth]{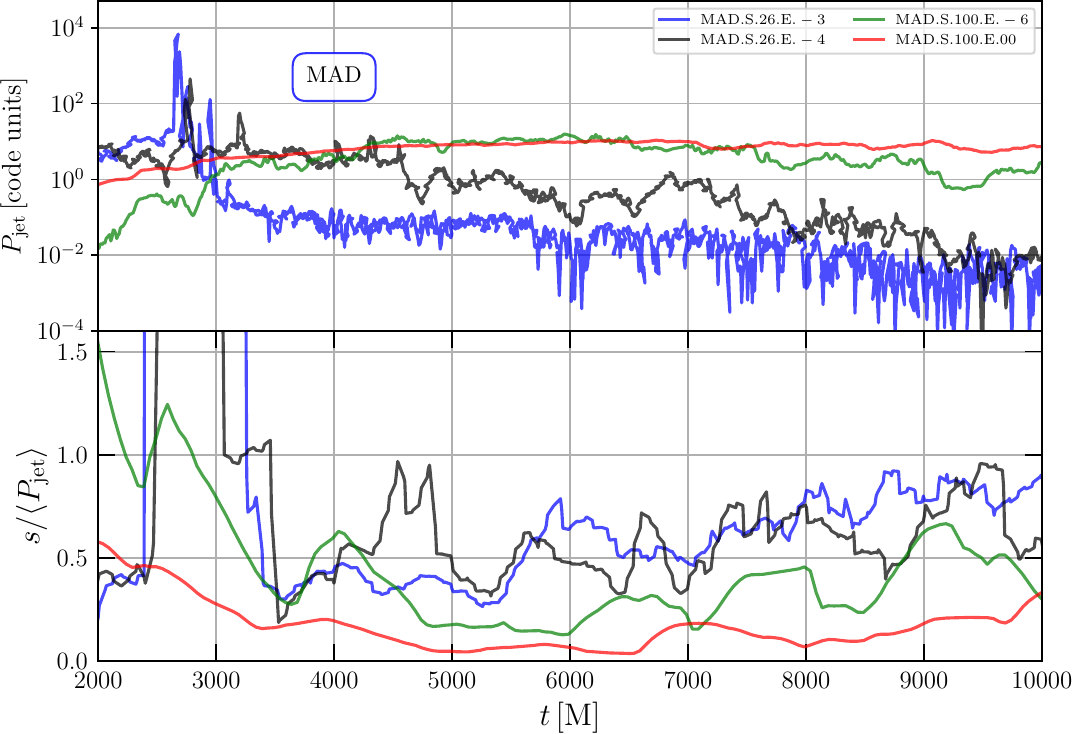}
  \end{center}
  \caption{ Upper panels: The power of the jet as defined in Eq. \eqref{pjet} in code units. Lower panel: the measure of variability for the jet power for four representative MAD models.}
    \label{fig:pow}
\end{figure}

The power of the jet of four MAD models is shown in the upper
panel of Fig. \ref{fig:pow}.
The ideal MAD model, ${\rm MAD.S.100.E.00}$, has a similar power output
with the low resistivity model, ${\rm MAD.S.100.E.-6}$. However, when 
the resistivity increases significantly the picture vastly changes.
For   models ${\rm MAD.S.26.E.-3}$ and ${\rm MAD.S.26.E.-4}$,
there is an initial jump at the normalized magnetic flux at
around $2500$ M (see Fig. \ref{fig:mdot-all}). A similar behavior is 
seen in the jet power, with an early burst followed by a steady 
decline in power throughout the simulation time. 

To measure the variability of the jet power we make use of 
Eq. \eqref{eqvar}, where $k_i=P_{\rm jet}$ in this case. The ideal 
model shows the smallest variability on the jet power, whereas 
all the rest resistive models, shown in the lower 
panel of Fig. \ref{fig:pow}, exhibit larger variability.

\section*{Appendix D. MRI quality factor $Q_\theta$}
\label{appD}

We provide here the definition of the MRI quality 
factor $Q_\theta$ and details on its calculation presented 
in Table~\ref{table:models}. We evaluate the so-called ``quality factor'' $Q_\theta$,
in terms of the ration between the grid spacing in a given direction
$\Delta x_{\theta}$, (\eg the $\theta$-direction) and the wavelength of
the fastest growing MRI mode in that direction (\ie $\lambda_{\theta}$),
where both quantities are evaluated in the tetrad basis of the fluid
frame $e_{\mu}^{(\hat{\alpha})}$ (see \citealt{Takahashi:2008,
  Siegel2013, Porth2019}, for details)
\begin{align}
  Q_{\theta}:=\frac{\lambda_{\theta}}{\Delta x_{\theta}}\,,
\label{Qth}
\end{align}
where
\begin{equation}
\lambda_{\theta}:=\frac{2\pi}{\sqrt{(\rho h +b^2)}\Omega}
b^{\mu}e_{\mu}^{(\theta)}\,,
\end{equation}
$\Omega:=u^{\phi}/u^t$ is the angular velocity of the fluid and the
corresponding grid resolution is $\Delta x_{\theta}:=\Delta
x^{\mu}e_{\mu}^{(\theta)}$. Finally the average of $Q_{\theta}$ is done in space and  time, specifically in a time window of 200 M and spatially in the region of interest at angles $60^o < \theta < 120^o$ and $r<40\, r_g$,
inside the heart of the disk.

\end{document}